\documentclass[intlimits,twoside,a4paper]{article}

\usepackage{amsmath,amssymb}
\usepackage{graphicx}
\usepackage[T2A]{fontenc}
\usepackage[cp1251]{inputenc}

\usepackage[eqsecnum]{cmpj2}

\usepackage[russian,english]{babel}
\usepackage{textcomp}
\usepackage{wrapfig}

\issue{2014}{17}{4}{43401}
\doinumber{10.5488/CMP.17.43401}

\title[Electric cell voltage at etching and deposition of metals under magnetic field]
{Electric cell voltage at etching and deposition \\
of metals under an inhomogeneous constant \\ magnetic field}
\author[O.Yu. Gorobets \textsl{et al.}]{O.Yu. Gorobets\refaddr{label1,label2}, Yu.I. Gorobets\refaddr{label1,label2}, V.P.~Rospotniuk\refaddr{label1}, Yu.A.~Legenkiy\refaddr{label3}}
\addresses{
\addr{label1} National  Technical University of Ukraine ``Kyiv Polytechnic Institute'', 37 Peremogy Ave.,  03056 Kyiv, Ukraine
\addr{label2} Institute of Magnetism of the National Academy of Sciences of Ukraine and the Ministry of Education and Science of Ukraine, 36-b Vernadsky Ave.,  03142 Kyiv, Ukraine
\addr{label3} Donetsk National University, 40 Krasnozorenska St., Donetsk-87, Ukraine}

\date{Received December 3, 2013, in final form June 16, 2014}
\authorcopyright{O.Yu. Gorobets, Yu.I. Gorobets, V.P.~Rospotniuk, Yu.A.~Legenkiy, 2014}

\begin{document}

\maketitle

\begin{abstract}
	The self-organized electric cell voltage of the physical circuit is calculated at etching and deposition of metals at the surface of a magnetized ferromagnetic electrode in an electrolyte without passing an external electrical current. This self-organized voltage arises due to the inhomogeneous distribution of concentration of the effectively dia- or paramagnetic cluster components of an electrolyte at the surface of a ferromagnetic electrode under the effect of inhomogeneous magnetostatic fields. The current density and Lorentz force are calculated in an electrolyte in the vicinity of the magnetized steel ball-shaped electrode. The Lorentz force causes the rotation of an electrolyte around the direction of an external magnetic field.
\pacs 41.20.Gz, 81.05.Zx, 81.65.Cf, 82.45.Bb, 82.45.Gj, 82.45.Hk
\end{abstract}

\vspace{1ex}

\section{Introduction}

Electrochemical deposition and etching of metals or electroplating are widely used in various fields of science and technology today. The rapid development of micro- and nanoelectronics, sensory systems, etc., has given impetus to the creation of new materials using the electroplating technology, providing high corrosion and wear resistance, permits to attach the surface of the desired shape and pattern as well as to improve the necessary electrical properties of the constituent circuits. The application of magnetic fields in electroplating could be an important step in solving the problem of creating thin films of variable thickness, high-quality masks and anisotropic etching of the appropriate areas of functional materials in micro- and nanoelectronics \cite{KuboApplPhys10, MatsushimaSolStElChem07, ChouchaneaAlloysComp10, TabakovicElChemSoc03}. However, the use of electrochemical deposition and etching of metals in homogeneous and inhomogeneous magnetic fields is not limited to the above examples and has a wide practical utilization.
	
The study of the kinetics of electrochemical processes in magnetic fields began
on the verge of %\linebreak
the 19th--20th centuries \cite{BagardJdePhys96, HeilbrunAnPhys04, HoraceJAmChemSoc13}.
However, due to the lack of theoretical understanding of these processes, the development of magnetoelectrolysis
as a field of science was slow enough. The rapid progress in the investigation of the magnetic field effect on the process of electrochemical deposition, etching and corrosion of metals, which can be seen from the analysis of work \cite{FahidyEncycl}, has taken place only for the recent 30--40 years. This progress is associated with the classical work by Fahidy \cite{FahidyApplElChem83}, which laid the theoretical foundations of the modelling of the effects of magnetoelectrolysis based on the combined system of the magnetohydrodynamic equations for a weakly conducting fluid and the convective diffusion equation. This system of equations describes a wide spectrum of magnetohydrodynamic (MHD) effects observed, when an electric current generated by external sources (i.e., in crossed electric and magnetic fields) flows through the electrolyte in a magnetic field \cite{TackenApplElChem95, KimElChemSoc95, JagannadhIndChemRes96, BundSciTechnol08}.
	
	In particular, it is shown in the paper \cite{FahidyElActa73} that the rate of mass transport increases due to the occurrence of convective diffusion flows under the effect of the Lorentz force and thus the rate of the etching and metal deposition processes increases. This effect, referred to as MHD effect, was confirmed in the papers \cite{AogakiDenki75, AogakiDenki76}. If the local current density within the diffusion layer near the electrode surface is not uniform due to the existence of the surface roughness, then the so-called micro-MHD effect is observed \cite{FahidyEncycl}. Manifestation of the micro-MHD effect on a much smaller scale [with the order of magnitude of the thickness of the diffusion layer ($\sim 10\div 100$~\textmu{}m)] accompanied by the formation of micro-vortexes was found in the study of copper electrodeposition in parallel magnetic and electric fields directed perpendicular to the electrode \cite{AsanumaApPhys05, AogakiJMMM10}. Thus, the research of the above mentioned electrochemical reactions and the mass transport in an uniform magnetic field \cite{FahidyElActa73} accompanying them laid the foundations for a new branch in electrochemistry, i.e.,  magnetoelectrochemistry.
	
	However, the gradient magnetic force acting on the ions in the electrolyte in an inhomogeneous magnetic field, in particular, created by a magnetized ferromagnetic electrode is not taken into account in the convective diffusion equation in the theory of magnetoelectrolysis of Fahidy \cite{FahidyApplElChem83}. This is due to the fact, that the gradient magnetic force acting on the ions is 5--6 orders of magnitude less than the so-called ``entropic'' force \cite{CoeyEurophysNews03} in the equation of convective diffusion. Therefore, the magnetic force acting on a unit volume of electrolyte (including the gradient magnetic force, gradient paramagnetic force and Lorentz force) is included only in the equation of magnetohydrodynamics \cite{FahidyApplElChem83, TackenApplElChem95, BundSciTechnol08}. At the same time, there are a number of experimental effects \cite{TangElSoc07, SueptitzElActa09, CostaJMMM04, PullinsChemB01, GorobetsMatSc07, GorobetsPhysChemC08, GorobetsPhysMet05, GorobetsApSurfSc05, GorobetsPhysSolC04, GorobetsMetalPhys06, SueptitzCorSc11, IlchenkoJMMM10, GorobetsPhysMet12, GorobetsFuncMat11, GorobetsVisnykDon09, DunnePhysRewB12, DunnePhysRewLett11, DunneMGD12, DunneApplPhys12, TschulikElCommun11, YangPhysChemLett12, UhlemannEurPhys13} that indicate the existence of magnetic gradient forces of considerable magnitude in the convective diffusion equation. Thus, the effect of an inhomogeneous magnetic field on corrosion, chemical etching and electrodeposition of metals is a subject of modern research in the field of magnetoelectrolysis. It was experimentally revealed in the papers \cite{TangElSoc07, SueptitzElActa09, CostaJMMM04, PullinsChemB01} that the peculiarities of the processes, mentioned above, depend on the value of the magnetic field gradient at the electrode surface. In particular, the ferromagnetic electrodes can create different distributions of an inhomogeneous magnetostatic field in an electrolyte depending on their shape, size and magnetization. Moreover, a  number of experimental effects are revealed that demonstrate a correlation of the etching pattern and spatial distribution of a magnetostatic field of magnetic domains for ferromagnetic samples \cite{GorobetsMatSc07, GorobetsPhysChemC08, GorobetsPhysMet05, GorobetsApSurfSc05, GorobetsPhysSolC04, GorobetsMetalPhys06, SueptitzCorSc11, IlchenkoJMMM10, GorobetsPhysMet12, GorobetsFuncMat11}. Such etching structures represent the repeating elevations and cavities of the sizes coinciding with the sizes of magnetic domains. The paper \cite{SueptitzCorSc11} is devoted to the analysis of corrosive areas that were experimentally revealed during the anodic dissolution of a permanent magnet NdFeB in sulfuric acid. It was also experimentally established \cite{TangElSoc07} using an iron electrode in the form of a cylindrical core, that the form of the etching figure and of the etched areas (chemical etching process was carried out in 1~M NaCl solution in the magnetic field of 0.35~T) strongly depends on the direction and magnitude of the applied gradient magnetic field. Thus, the size and depth of the pits formed during the etching process in the magnetic field directed along the main axis of the cylindrical sample are much bigger than the characteristic scales of these regions formed after the etching without applying an external magnetic field. Similarly, in the reference \cite{IlchenkoJMMM10}, the anisotropic etching of a magnetized steel ball is revealed having an obvious elongation of the etching figure along the direction of an external magnetic field. The effects of the dependence of a deposit structure on the spatial distribution of the magnetostatic field at the electrode surface have been also observed at metal deposition from the appropriate electrolyte solutions \cite{GorobetsPhysChemC08, GorobetsPhysMet12}. In particular, the periodic structure of the dendrite regions is created at nickel electrodeposition under spatially nonuniform magnetostatic field, while the dendritic regions are localized at the regions of the cathode with maximum magnetic field strength \cite{GorobetsPhysChemC08}. It is also noted \cite{DunnePhysRewB12, DunnePhysRewLett11, DunneMGD12, DunneApplPhys12, TschulikElCommun11, YangPhysChemLett12, UhlemannEurPhys13} that the thickness of the diffusion layer (at a distance $\sim 100$~\textmu{}m from the electrode surface) changes by changing the applied magnetic field in the process of electrodeposition of paramagnetic cations from the solution onto the surface of the electrode in the form of plates. The inhomogeneous magnetic field, created using a magnetic lattice, forms periodic dotted hexagonal \cite{DunnePhysRewB12, DunnePhysRewLett11, DunneMGD12, DunneApplPhys12} or circular \cite{DunnePhysRewB12, DunnePhysRewLett11, DunneMGD12, DunneApplPhys12, TschulikElCommun11, YangPhysChemLett12, UhlemannEurPhys13} structures of the deposition areas. The structures of the sediment slightly change their size and location depending on the magnitude and the direction of gradient magnetic field and the type of dissolved ions. Convection also directly affects the dynamics and the shape of deposition by blurring the sediment depending on the mutual direction of the external magnetic field and the electrode in the gravity field. In a number of papers \cite{TschulikElCommun11, YangPhysChemLett12, UhlemannEurPhys13} it is shown that under certain conditions, the electrochemically active ions Cu$^{2+}$, Fe$^{2+}$, Co$^{2+}$ formed the thickest layer of the sediment near the areas with the maximum gradient of the magnetic field. However, under other conditions the electrochemically active ions Bi$^{3+}$ and Cu$^{2+}$  form the sediment localized at the electrode surface, where the magnetic field gradient is minimal, in contrast to the previous case. The last type of the anisotropy of the sediment is observed after adding electrochemically inert ions Mn$^{2+}$ to the solution. Inverse regions were formed in the process of deposition of the ions Zn$^{2+}$ and Cu$^{+}$, but with the addition of electrochemically inactive ions Dy$^{3+}$, Gd$^{3+}$ to a solution \cite{DunnePhysRewB12, DunnePhysRewLett11, DunneMGD12, DunneApplPhys12}.

	Furthermore, the fluid movement is observed far beyond the diffusion layer of a ferromagnetic electrode in the processes of deposition, etching and corrosion of metals in an external magnetic field without an external electric current passing through an electrolyte solution \cite{IlchenkoJMMM10, GorobetsFuncMat11, GorobetsVisnykDon09, GorobetsMHD03}. The phase separation of an electrolyte of the ``liquid--liquid'' type with the sizes of phase areas characterized by the same typical scale as the size of the electrode, is also observed in an inhomogeneous magnetic field created by the magnetized ferromagnetic electrode \cite{GorobetsVisnykDon09}.

	As it was already mentioned, the expected effect of the inhomogeneous magnetic field of moderate strength ($<10$~kOe) on diffusion processes, stationary distribution of concentration of paramagnetic ions and electrochemical reaction rate should be negligible without an external electric current passing through an electrolyte. This is connected with quantitative estimations of the energy of a separate paramagnetic ion at room temperatures in an external magnetic field of a moderate strength taking into account its magnetic susceptibility. The magnetic energy of a paramagnetic ion is at least $4\div 6$ orders of magnitude less than ${{k}_{\mathrm{B}}}T$, where ${{k}_{\mathrm{B}}}$ is Boltzmann constant, $T$ is the absolute temperature. This again illustrates the negligible effect of magnetic forces in comparison with the entropic force in the convective diffusion equation in view of Einstein relation between the diffusion coefficient and the mobility~\cite{Levich62}.

	In this regard, this work is devoted to creating a physical model of the effect of gradient magnetic forces directly on the diffusion in the electrolyte and on the generation of a nonuniform distribution of the electric potential on the electrode surface in the above processes of magnetoelectrolysis without an external electric current passing through the electrolyte. The proposed model assumes that the processes of etching and deposition of metals on the surface of a magnetized electrode are accompanied by the formation of cluster components of an electrolyte, which are products of electrochemical reactions occurring at the electrode surface in the electrolyte under a magnetic field. These clusters are characterized by an average magnetic moment in a magnetic field which is by several orders of magnitude larger than the average magnetic moment of a single paramagnetic ion. These cluster products of electrochemical transformations may represent the paramagnetic or effectively paramagnetic coacervates \cite{JongKolloid30, BamfordTrFaraday50, OverbeekCellCompPhys57, RozenblatMicroencaps89}, bubbles of gases (including nanobubbles \cite{MatsushimaSolSt12, FanJinPhysChemB07, AttardAdvColl03, AttardPhysA02, AgarwalChem11,  ZhangLangmuir08}), nano- and/or microparticles with their ionic environment~\cite{TyrrellPhysRewLet01}.

	In this case, the estimation of the ratio of the potential magnetic energy of nanocluster components of an electrolyte with volume $V$ to the average kinetic energy of their thermal motion $\varepsilon =\chi V\vec{H}^2/2k_{\mathrm{B}}T$ is not negligible in the external magnetic field or in the magnetic field of the magnetized ferromagnetic electrode ($H\leqslant 10$~kOe), and thus the magnetic forces cannot be neglected in the diffusion equation in comparison with the entropic force. The previous estimation is provided for cluster components with typical effective magnetic susceptibility per unit volume about $\chi \cong {{10}^{-5}}\div {{10}^{-4}}$. The effective magnetic susceptibility is the difference between the magnetic susceptibility of the cluster component and the magnetic susceptibility of the electrolyte. It is easy to see that the thermal motion of the appropriate size (above $10\div100$~nm depending on their effective magnetic susceptibility) cluster products of chemical reactions does not prevent their magnetic capture.

	In this paper, the peculiarities of the processes of chemical etching and deposition of metals are considered under the effect of inhomogeneous constant magnetic fields by the example of a ball-shaped ferromagnetic electrode, magnetized in an external homogeneous magnetic field. Such a choice of the electrode shape is connected with the equivalent state of all the points at the surface of the ball in the absence of its magnetization. That is why it is easy to separate the magnetic field effects in such a system from the effects of another nature.
	\par	
	In particular, the expressions will be obtained for the electric cell voltage of the concentration circuit (that arises in the processes of chemical etching and deposition of metals in an electrolyte at the surface of ferromagnetic electrode which has its own inhomogeneous magnetic stray field caused by magnetization of the electrode in an external magnetic field), the electric current density and the Lorentz force. Comparison of experimental observations of the features of the dynamics of the described processes and the calculation results will be represented for a ferromagnetic ball-shaped electrode. The characteristic distance of the decay of gradient magnetic forces created by magnetized ferromagnetic electrodes on clusters in the electrolyte matches the order of the characteristic size of the electrode. Therefore, the obtained theoretical expressions are applicable at distances of the order of typical sizes of the electrode, which is much greater than the thickness of the diffusion layer, i.e., at the macro-scales, on which the above effects are observed.

	It is well known that there are two types of electromotive forces in electrochemistry between the electrodes (or surface regions of the same electrode) of the same chemical composition. The first type of electromotive forces is called the physical circuit. The source of the electric cell voltage in a physical circuit is the difference of the Gibbs energy which is caused by the difference of the physical states of the electrodes of the same chemical composition \cite{Bard01, Buchanan00, Vetter61, Thirsk74, Antropov72}. These electrodes are immersed in the same solution, and the electrode that has a less stable state transforms to a more stable state while the circuit operates. The typical example of the physical circuit is the gravitational one when the electrode of the bigger height dissolves because it has a bigger Gibbs energy in comparison with the shorter electrode. The cumulative process takes place till the height of the electrodes becomes the same \cite{Vetter61, Thirsk74, Antropov72}.

	 The other type of the electromotive force is the so-called concentration circuit where the electric cell voltage arises due to the difference of ion concentration in an electrolyte at the surface of the electrodes of the same chemical composition. The direction of the electrical current in an electrolyte promotes equalization of the concentrations mentioned above for a concentration circuit \cite{Vetter61, Thirsk74, Antropov72}.

	In general, the difference in physical properties of the electrode of an arbitrary shape and magnetization distribution as well as the concentration effects can appear in the experimental observation of the effect of inhomogeneous magnetic fields at the electrode surface on the processes of corrosion, chemical etching and electrodeposition. The choice of the ball-shaped ferromagnetic electrode to illustrate the model proposed in this paper is not occasional because it makes it possible to separate the effects of the first and the second types. Indeed \cite{KittelPhysRew48}, the vectors of magnetic field strength and magnetic field flux density are distributed uniformly inside the ferromagnetic sample at its magnetization for the ferromagnetic sample in the form of a three-axial ellipsoid (in particular, ferromagnetic ellipsoid of revolution, ball, cylinder and plate). That is why the magnetic part of the chemical potential and the corresponding physical state of atoms is the same and not dependent on the coordinates inside the uniformly magnetized ferromagnetic ball. The inhomogeneous distribution of magnetostatic fields arises outside the magnetized steel ball leading to the creation of inhomogeneous distribution of concentration of paramagnetic electrolyte components at its surface. Let us note that the approach taken in our work is not limited to the example of the chosen geometry of the shape of the electrode and is general in nature for the uniformly magnetized three-axial ellipsoid.

\section{Experimental methods }

\begin{wrapfigure}{i}{0.5\textwidth}
%\begin{figure}[htb]
\vspace{-5mm}
\centerline{\includegraphics[width=0.45\textwidth]{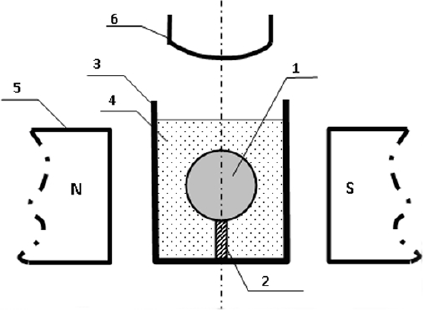}}
\caption{The scheme of the experimental setup.} \label{figScheme}
%\end{figure}
\end{wrapfigure}
The experiments of this work are conducted to compare the peculiarities of the processes of chemical etching and deposition of metals at the surface of a magnetized ferromagnetic electrode with the results of theoretical modelling. The installation and experimental procedure was used similarly to the one described, for example, in the paper \cite{IlchenkoJMMM10}. The experimental setup is represented in figure~\ref{figScheme}.
	
The steel ball (1) is fixed at dielectric holder (2) at the center of glass container (3) inside the working volume of magnetic system (5). The external magnetic field was directed horizontally. An external electric field was not applied. The container was filled with an electrolyte solution (4), in which the chemical dyes K$_{3}[$Fe(CN)$_6]$ or AgNO$_3$, were added for visualization of Fe$^{2+}$ ions. The observation was carried out by means of an optical microscope (6). A number of experiments were carried out with a thin uniform layer of another metal preliminarily deposited at the steel ball surface having the thickness much less than the ball radius.

\begin{wrapfigure}{i}{0.5\textwidth}	
%	\begin{figure}[htb]
\centerline{\includegraphics[width=0.47\textwidth]{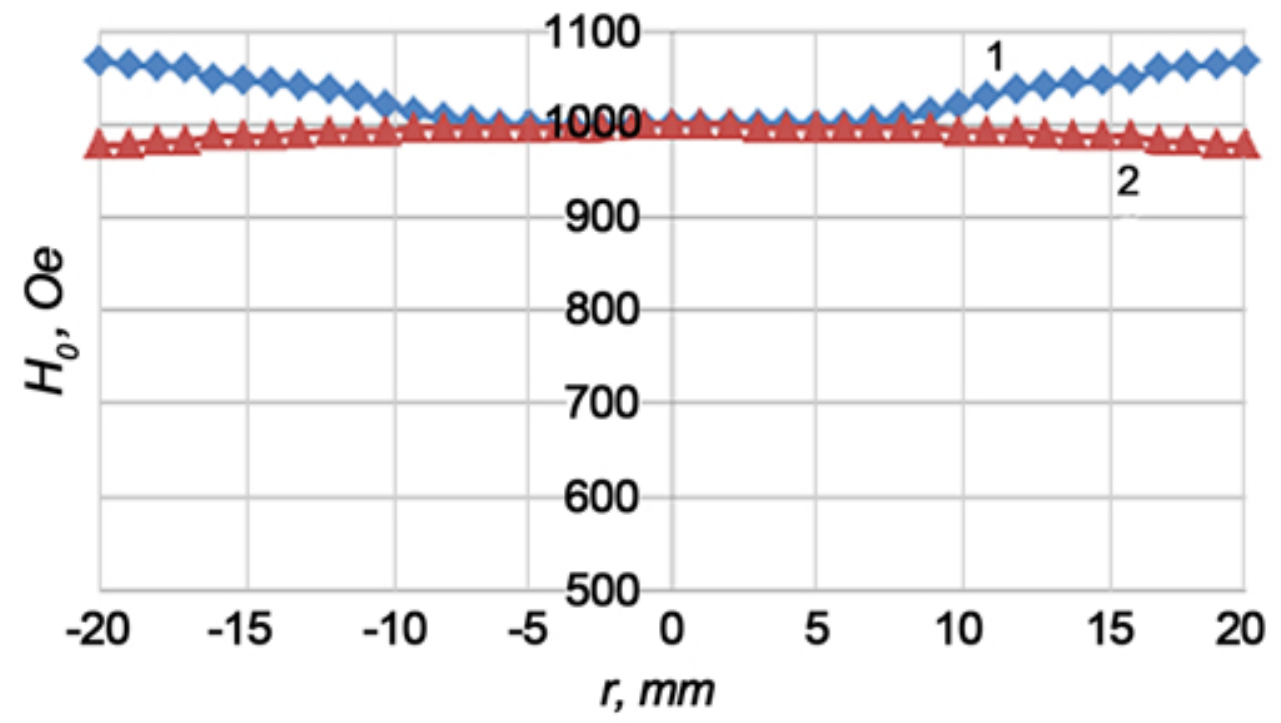}}
\caption{(Color online) Distribution of the magnetic field in the central plane of the space between the poles of the magnetic system in parallel (curve 1) and perpendicular (curve 2) directions to magnetic field lines. Measurement error is 1\%.} \label{figDistrMF}
%\end{figure}
\end{wrapfigure}
The magnetic system represents an electromagnet. The poles of the core of magnetic system have a diameter of 80~mm. The distance between the poles of the magnet was 50~mm in these experiments. In experiments with the balls of 2~mm radius, the cuvette represents a rectangular glass box having the following internal dimensions: length~--- 30~mm, width~--- 17~mm, height~--- 34~mm. The external magnetic field was directed horizontally. Heterogeneity of the magnetic field in the central part of magnetic system, on which the steel ball was placed, does not exceed the value of 1~Oe/mm in the parallel and perpendicular directions to the magnetic field without a steel ball in the system. Figure~\ref{figDistrMF} presents data on the distribution of the magnetic field along two lines that lie in parallel (curve 1) and perpendicular (curve 2) directions to magnetic field lines in a horizontal plane and pass through the center space between magnet poles. The distance from the center of the space between the poles of the magnet, where the center of the steel ball was placed in experiments, is measured on the horizontal axis. The magnitude of the magnetic field in oersteds is measured on the vertical axis.

{All studies were conducted using samples of a spherical shape and made of steel of the grade \textcyrillic{ШХ}--15 (\textcyrillic{ГОСТ}~801--78) (hereinafter referred to as steel balls).}

	The surface of the samples was prepared before the experiments in several stages. The lubricant and other contaminants were removed by mechanical means (with the help of a wiping cloth). Then, objects were degreased in a saturated solution of Noah
for 3 minutes, then extracted from the alkali solution and thoroughly washed in distilled water. Thereafter, the objects were dried and pickled in a solution of nitric acid with a concentration of 2.5\% within 1.5 minutes. Then, the objects were again washed with distilled water and dried.

The aqueous solutions of nitric, hydrochloric and sulphuric acids were used as electrolytes to investigate the processes of corrosion and chemical etching.

	Solutions for cementation of copper deposition were prepared from copper sulphate, 5-water qualification of ``W''.

Digital photo- and video shooting of the heterogeneous state of an electrolyte was carried out
during the experiment. This applies to both cases, i.e., without adding chemical dyes
to the electrolyte and with their use in order to increase contrast. Let us note that
the use of chemical dyes to visualize the distribution
of concentration of cluster
components of an electrolyte did not lead to noticeable changes of the observed processes.
Measurements of geometrical parameters of the deposited layers and phase regions of
the solution were carried out using computer processing of photo and video materials.

\section{Results}

\begin{wrapfigure}{i}{0.5\textwidth}
%	\begin{figure}[tb]
\centerline{\includegraphics[width=0.40\textwidth]{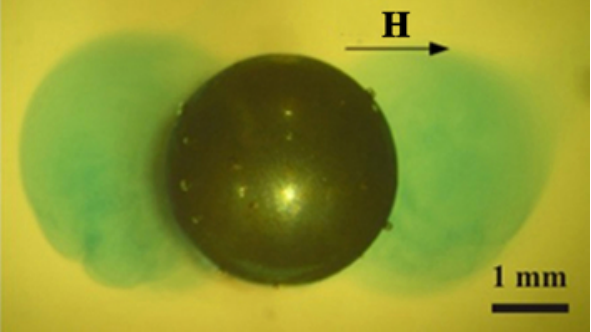}}
\caption{(Color online) A steel ball with the initial radius $R=1.5$~mm in the 2.5\% solution of hydrochloric acid with a chemical dye added for visualization of Fe$^{2+}$ ions in an external uniform magnetic field $H=3500$~Oe.}
\label{figBlueReg}
%	\end{figure}
\end{wrapfigure}

At the beginning of etching, a blue boundary arises around the magnetized steel ball as shown in figure~\ref{figBlueReg}. Further, the picture, similar to the one represented in figure~\ref{figBlueReg}, is formed at low acid concentrations ($<$~2.5\%).
	
Two blue regions are formed in the inhomogeneous magnetic field of a magnetized ball (figure~\ref{figBlueReg}) which confirms an increase of concentration of Fe$^{2+}$ ions in these regions.
	
The shape and localization of the regions of elevated concentration of  paramagnetic corrosion products (figure~\ref{figBlueReg}) are similar to the clusters of paramagnetic particles trapped in an inhomogeneous magnetic field of a magnetized steel ball \cite{FriedlaenderIEEE81, FriedlaenderIEEE82}.

It is possible to conclude on the basis of figure~\ref{figBlueReg} that the regions of elevated
concentration of paramagnetic Fe$^{2+}$ ions are formed in an electrolyte in the vicinity of
magnetic poles of a magnetized steel ball. The Fe$^{2+}$ ions get into an electrolyte at dissolution
of the steel ball surface. Moreover, the gradient magnetic field of a steel ball redistributes the
paramagnetic corrosion product concentration in its vicinity due to the forces connected with the
gradient of the square of magnetic field strength and can effect the chemical processes at its surface.

\begin{figure}[!h]
\centerline{\includegraphics[width=0.6\textwidth]{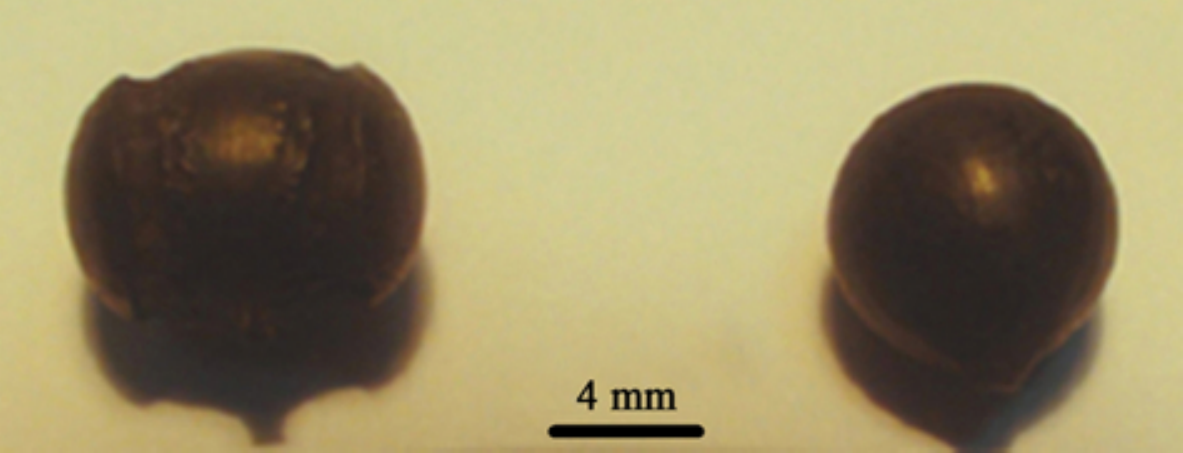}}
\caption{(Color online) Optical image of a sample after 1.5~hours of etching of a steel ball with the initial diameter of 8~mm in the 15\% solution of nitric acid: under the 3~koel magnetic field (left-hand) and without application of a magnetic field (right-hand). The place of fixing a steel ball to the holder is visible at the etching figures both with and without application of a magnetic field.}
\label{figEtching}
	\end{figure}

	As it is shown in the reference \cite{IlchenkoJMMM10}, the an\-isotropy of the etching figure
is observed at steel ball etching in a magnetic field in the course of time. The obvious elongation
of the etching figure along the direction of an external magnetic field is observed, for example,
as it is shown in figure~\ref{figEtching}.

\begin{figure}[!b]
\centerline{\includegraphics[width=0.5\textwidth]{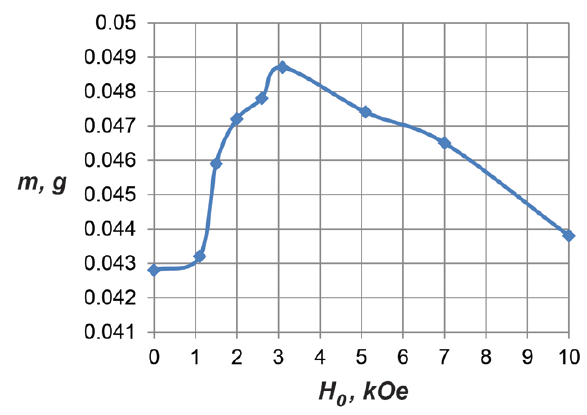}}
\caption{Dependence of sample mass for the steel ball of 2.5~mm initial diameter on an external magnetic field strength after 20~min etching in 15\% nitric acid solution. Measurement error is 4\%.}
\label{figEtchedMass}
	\end{figure}

	The experimental dependence of the sample mass on the magnetic field strength is presented in figure~\ref{figEtchedMass} for the steel ball of 2.5~mm initial diameter and 15\% nitric acid concentration.

\begin{figure}[!t]
\centerline{\includegraphics[width=0.5\textwidth]{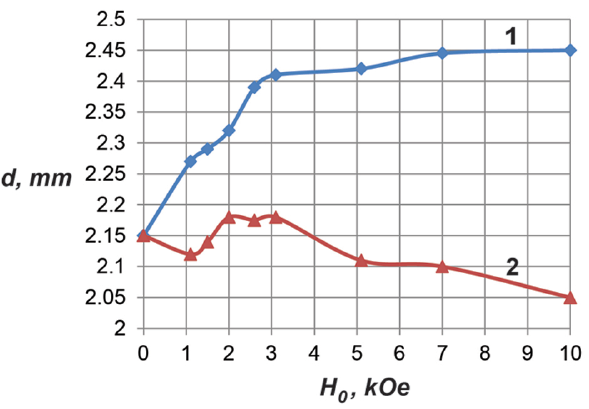}}
\caption{(Color online) Dependencies of ball diameters at pole and at equator on an external magnetic field strength for the steel ball of 2.5~mm initial diameter and 15\% HON$_3$ concentration after 20~min etching. Curve 1 presents the dependence of ball diameters at pole on an external magnetic field, and curve 2 presents the dependence of ball diameters at equator on an external magnetic field. Measurement error is 1.5\%.}
\label{figEtchedDiameter}
	\end{figure}

	Dependencies of ball diameters at pole and at equator on an external magnetic field strength are presented in figure~\ref{figEtchedDiameter} for balls of 2.5~mm initial diameter and 15\% HON$_3$ concentration after 20~min etching.

	The aqueous solutions of  CuSO$_4$ and AgNO$_3$ have been used for the investigation of the processes of metal deposition at the surface of a magnetized steel ball. The contact exchange reaction at the displacement of copper from copper sulphate solution by iron is the dominant chemical reaction during the process of deposition of copper onto the surface of the iron ball. It is described by the following equation:
	\begin{equation*}	
	\text{Fe}^0+\text{Cu}^{2+}=\text{Fe}^{2+}+\text{Cu}^0,
	\end{equation*}	
whence it is easy to see that  ion Fe$^{2+}$ is one of the ions in the solution at the outlet of the reaction.
	
\begin{figure}[!b]
\centerline{\includegraphics[width=0.4\textwidth]{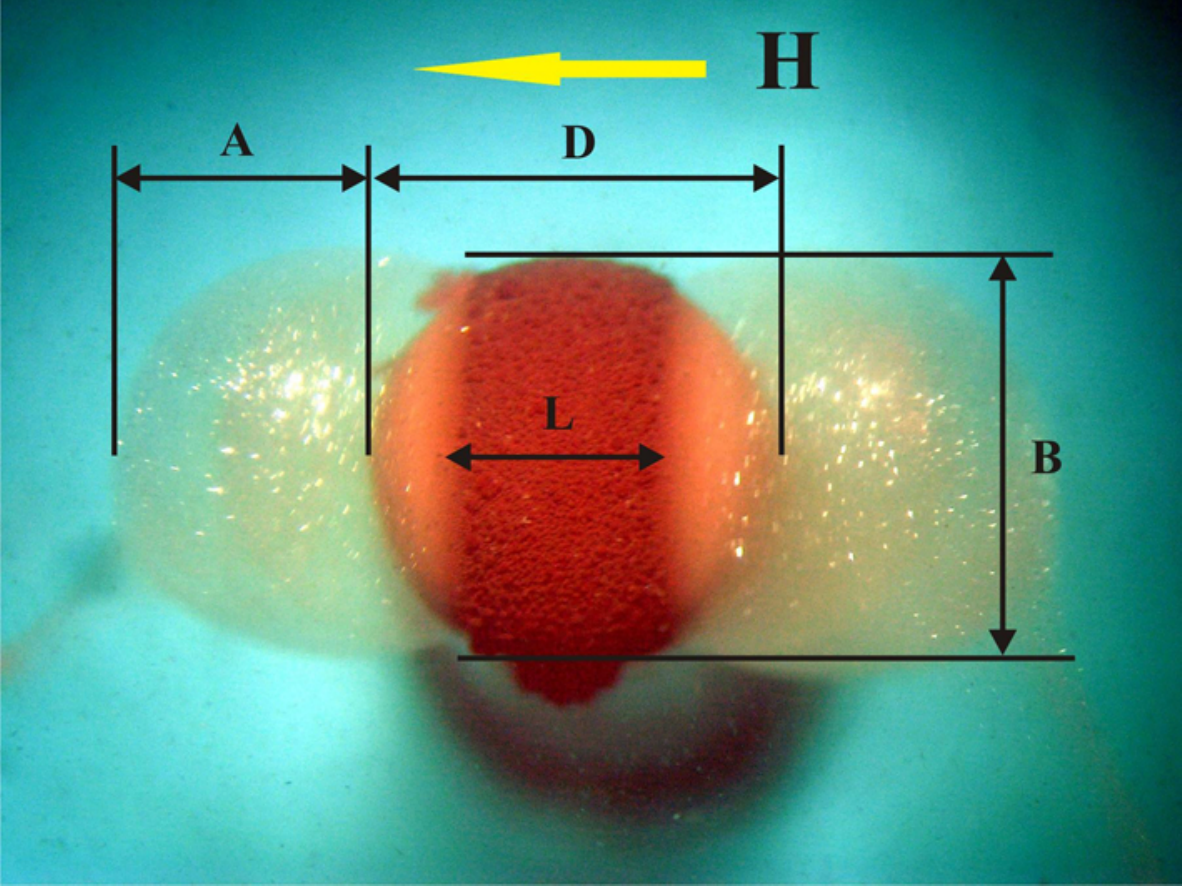}}
\caption{(Color online) Steel ball with the initial radius of $R=2$~mm in the CuSO$_4$ solution with
addition of the chemical dye for visualization of Fe$^{2+}$ ions (solution of AgNO$_3$) in the uniform
horizontal external magnetic field $H=1000$~Oe.}
\label{figVisual}
\end{figure}

\begin{figure}[!t]
\centerline{\includegraphics[width=0.65\textwidth]{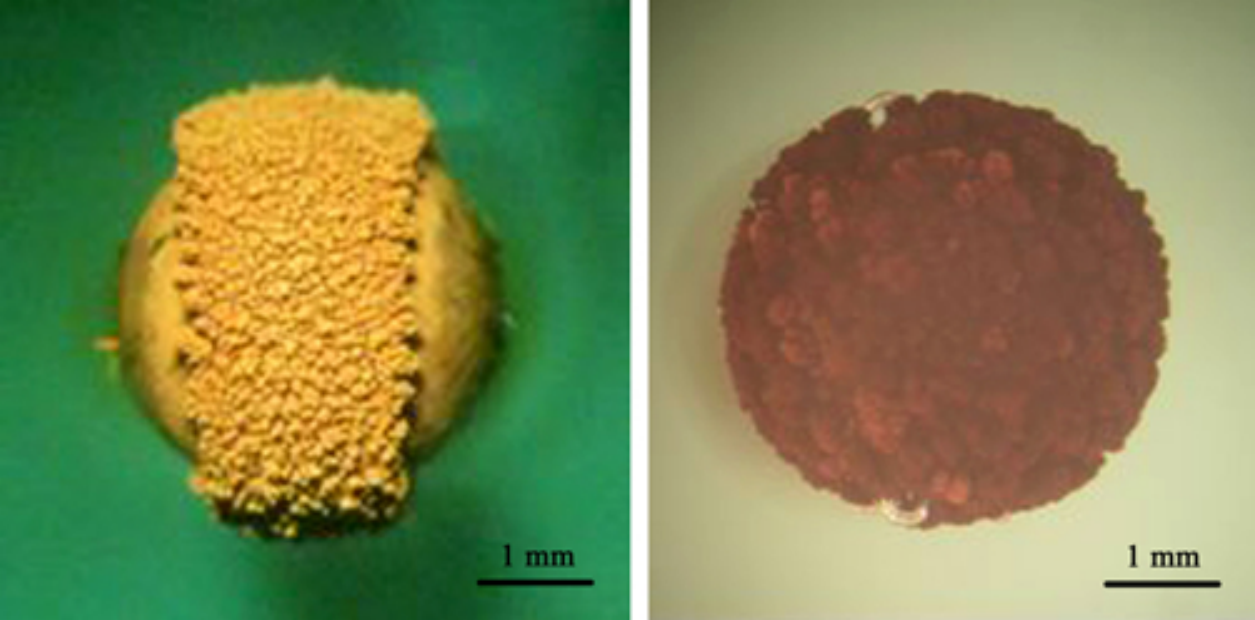}}
\caption{(Color online) Optical image of copper deposit from the CuSO$_4$ solution at the
steel ball with the initial radius of $R=2$~mm in the uniform horizontal external magnetic
field $H=1000$~Oe (left-hand) and without a magnetic field (right-hand). Duration of
deposition is about 90~minutes.}
\label{figCopperDepBall}
	\end{figure}

	Two regions of elevated concentrations of Fe$^{2+}$ ions are also formed in the vicinity of the magnetic poles of a magnetized steel ball at copper deposition figure~\ref{figVisual}. The anisotropy of the copper deposit is also observed relative to the direction of an external magnetic field figure~\ref{figCopperDepBall}.

	In order to visualize the localization zone of clusters containing iron ions (figure~\ref{figVisual}),
a weak 1\%-solution AgNO$_3$ containing the ions Ag$^+$ was added to a solution of CuSO$_4$.
It is well known that if ions Fe$^{2+}$ and Ag$^+$ are simultaneously present in the solution,
then the reaction with the formation of fine particles of metallic silver is possible:
	\begin{equation*}	
	\text{Fe}^{2+}+\text{Ag}^+=\text{Fe}^{3+}+\text{Ag}^0.
	\end{equation*}	
However, the last reaction is not dominant in this process.

	Similar chemical reactions can be written for the cases of other transformations discussed above \cite{Vetter61}.
	
\begin{figure}[!b]
\vspace{1ex}
\centerline{
\includegraphics[width=0.49\textwidth]{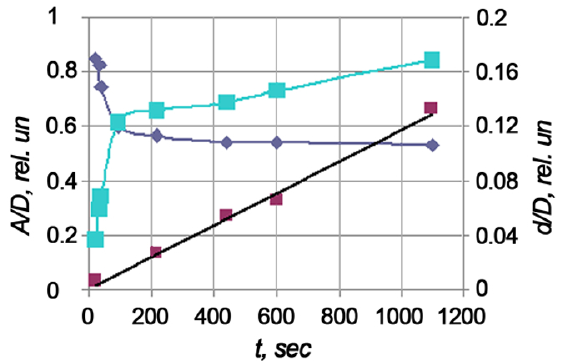}
\hspace{10mm}
\includegraphics[width=0.37\textwidth]{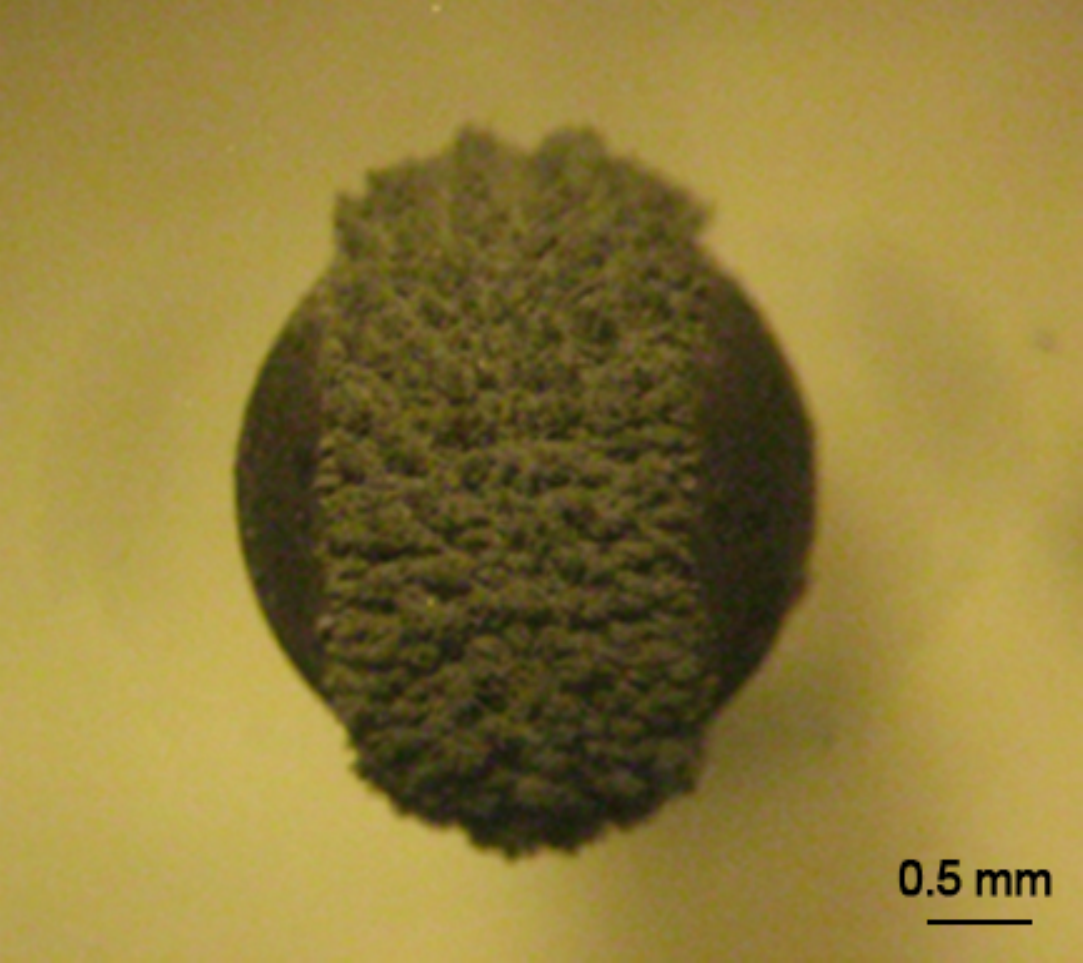}\hspace{5mm}
}
\parbox[t]{0.5\textwidth}{
\caption{(Color online) Dependencies of the parameters $A$, $L$ and $d$ on the time of deposition. Blue squares correspond to $A$, dark blue diamonds~--- $L$, red squares~--- $d$. Measurement error is~2\%.}
\label{figDepParameters}
%\end{figure}
%\end{wrapfigure}
%\begin{figure}[!t]
}
\parbox[t]{0.5\textwidth}{
\caption{(Color online) Optic image of the silver deposit from the AgNO$_3$ solution at the steel ball with the initial radius of $R=1.5$~mm in the uniform horizontal external magnetic field $H=1000$~Oe. Duration of deposition is about 10~minutes.}
\label{figSilverDepBall}
}
\end{figure}

	In figure~\ref{figVisual}, the parameter $A$ characterizes the size of the region with a higher
concentration of ions Fe$^{2+}$ along the magnetic field direction, parameter $L$ is the width of
deposition of copper dendrites, parameter $B$ is proportional to the thickness of the deposited
layer of copper dendrites at the equator (figure~\ref{figVisual}, figure~\ref{figCopperDepBall}),
$D=2R$ is the apparent diameter of the ball. Measurements of these parameters were performed
in the paper depending on the deposition time. Figure~\ref{figDepParameters} shows the results of
these measurements for the ball of radius 2~mm at the cementation of copper deposition
in the magnetic field of $H=2500$~Oe, that are normalized to the size of the diameter $D$ of the ball.	
	
%\begin{wrapfigure}{i}{0.5\textwidth}

In figure~\ref{figDepParameters}, parameter $d$ characterizes the thickness of the deposited layer of copper dendrites at the equator and is calculated as $d=B-D$ (see figure~\ref{figEtching}). It is seen from figure~\ref{figDepParameters} that the width of the deposited copper layer correlates with $L$~--- the size of the region with a higher concentration of ions Fe$^{2+}$. The thickness of the layer of copper dendrites at the equator increases monotonously with the deposition time.

	The magnetic field effects of silver deposition at the steel ball surface (figure~\ref{figSilverDepBall}) are similar to the effects for copper deposition described above. It is worthy to mention that the silver deposit structure is spherically symmetric without magnetic field application as well as for the case of copper.

	The processes of copper and silver deposition from the CuSO$_4$ and AgNO$_3$ solutions have been also investigated for the case when the thin uniform layer of zinc was preliminarily deposited at the steel ball surface with the thickness much less than the ball radius.

\begin{figure}[!t]
\centerline{\includegraphics[width=0.65\textwidth]{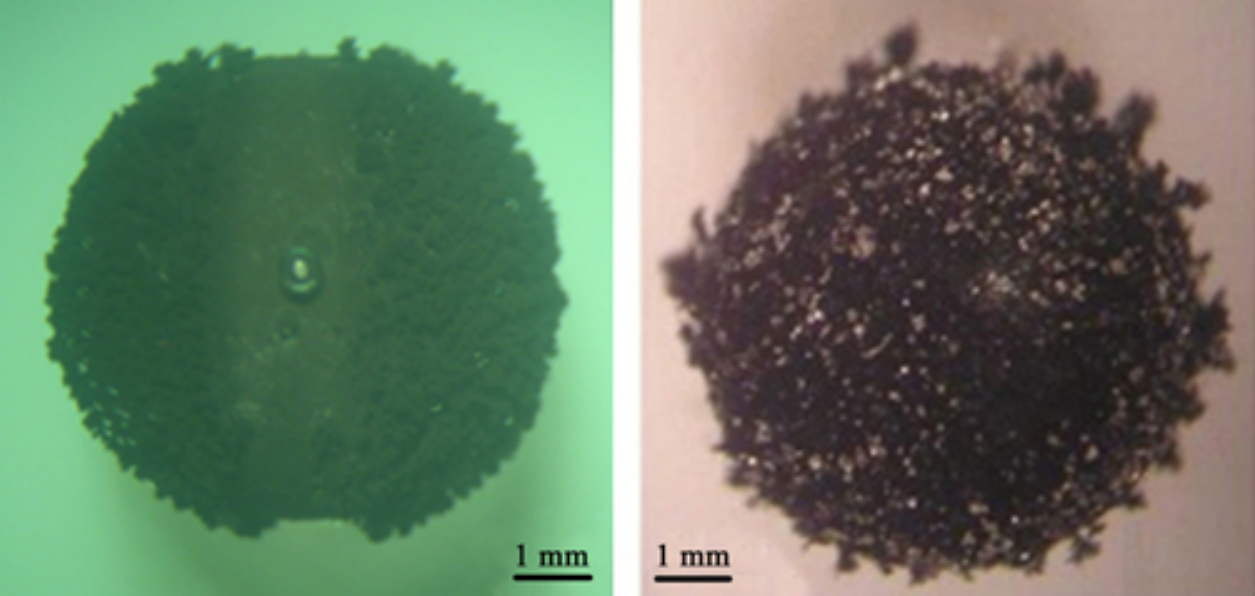}}
\caption{(Color online) Optic image of copper deposit (left-hand) from the CuSO$_4$ solution and silver deposit (right-hand) from the AgNO$_3$ solution at the zinc-plated surface of a steel ball with the initial radius of $R=2$~mm in the uniform horizontal external magnetic field $H=1000$~Oe.}
\label{figDepBallZinc}
	\end{figure}

	Experiments have shown that the copper deposit anisotropy is observed relative to the direction of an external magnetic field (figure~\ref{figDepBallZinc}, left-hand) at copper deposition at the surface of the magnetized zinc-coated ferromagnetic ball. However, the elongation (figure~\ref{figDepBallZinc}, left-hand) of the copper deposit is characterized by the opposite direction relative to the external magnetic field direction in comparison with copper deposition at non zinc-plated steel ball surface.

	There is no silver deposit anisotropy relative to the external magnetic field direction at silver deposition at the zinc-coated surface of a magnetized ferromagnetic ball (figure~\ref{figDepBallZinc}, right-hand) in contrast to the silver deposition at the steel surface of a magnetized ball (figure~\ref{figSilverDepBall}).

	However, the above mentioned experiments, can be explained exclusively on the basis of the model of continuous medium which was elaborated earlier and had a wide approbation for modelling the high gradient magnetic separation of weakly magnetic particles having magnetic susceptibility different from the magnetic susceptibility of a working fluid \cite{FriedlaenderIEEE81, FriedlaenderIEEE82}.

	Moreover, it is the effective magnetic susceptibility that ``works'' in the model of a continuous medium, i.e., in thermodynamic approximation. Consequently, the phase with bigger magnetic susceptibility displaces the phase with lower magnetic susceptibility into the region of minimum magnetic field strength.  While applying the model of the continuum for the description of the experiments of this work it is supposed that a certain part of dia- or paramagnetic ions move in the solution not separately from each other but in the ``bounded state'' as a constituent part of groups (i.e., clusters) of magnetic ions~--- further referred to as magnions. In this model, the magnetic energy of a group of paramagnetic ions can be proportional to the quantity of the ions in the group $n_{\mathrm{g}}$. That is why the concentration difference of magnions at the poles and at the equator of the magnetized ball can be great as it is observed experimentally.

	Let us consider magnetic properties of Fe$^{2+}$, Cu$^{2+}$, Ag$^+$, Zn$^{2+}$ ions that were used in the experiments of this work for qualitative explanation of anisotropy of the electrochemical reaction rates under inhomogeneous magnetic field at the electrode surface. Thus, the Fe$^{2+}$ paramagnetic ions possess a greater magnetic moment of $\mu_{\text{Fe}^{2+}} =5.4\mu_0$ (where $\mu_0$ is Bohr magneton) than the paramagnetic Cu$^{2+}$ ions ($\mu_{\text{Cu}^{2+}} =1.9\mu_0$) or diamagnetic Ag$^+$ and Zn$^{2+}$ ions \cite{Krynchik76}.

	Thus, the Fe$^{2+}$-containing magnions move the pole regions with a maximum magnetic field flux density under the effect of the gradient magnetic force at copper and silver deposition at the steel ball and at etching a steel ball in diamagnetic solutions of acids. They displace the solutions of copper or silver ions, or acid solutions from the pole regions as less magnetic. The regions of elevated concentration of the Fe$^{2+}$-containing magnions are formed near the poles of the ball (figure~\ref{figVisual}). The effective magnetic susceptibility of the Fe$^{2+}$-containing magnions is greater than zero in these cases. Maximum copper and silver deposition rates are observed at the magnetic equator as well as maximum steel dissolution rate is observed correspondingly at the conditions of clearly diffusive kinetics due to a greater rate of disposal of the reaction product (Fe$^{2+}$-containing magnions in this case) at the equator. The Zn$^{2+}$-containing magnions are effectively diamagnetic at copper deposition at the zinc-plated surface of a magnetized steel ball in the next experiment (figure~\ref{figDepBallZinc}, left-hand). Reaction product-containing magnions have a major effect on the reaction kinetics because they are accumulated due to the gradient magnetic force.
It is worth noting that the absolute value of the magnetic susceptibility of diamagnetic ions is approximately two orders of magnitude less than the magnetic susceptibility of paramagnetic ions. The same relation is valid for the gradient magnetic force and the magnetic energy acting at magnions containing diamagnetic ions in a diamagnetic solution. That is why there is no anisotropy of deposit in the experiments of silver deposition at the surface of a zinc-plated magnetized ball from the diamagnetic AgNO$_3$ solution (figure~\ref{figDepBallZinc}, right-hand). Gradient magnetic forces are negligible under such experimental conditions at moderate external magnetic fields.

\begin{wrapfigure}{i}{0.5\textwidth}	
%\begin{figure}[htb]
\centerline{\includegraphics[width=0.45\textwidth]{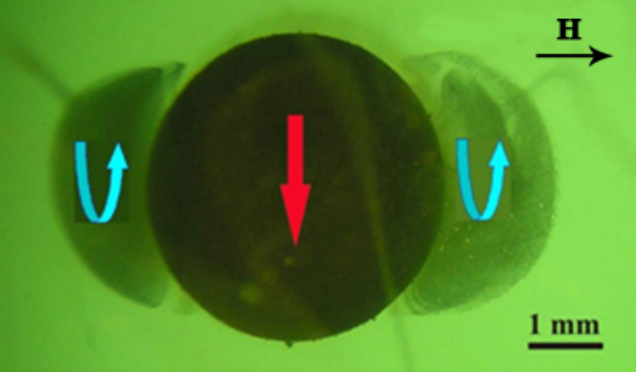}}
\caption{(Color online) Scheme of electrolyte rotation in the processes of corrosion and deposition at the surface of a steel ball in a horizontal magnetic field.}
\label{figElRotat}
%\end{figure}
\end{wrapfigure}
There is another experimental fact which has the opposite manifestation for the processes of copper deposition at the steel and zinc-plated surface of a magnetized steel ball in a magnetic field. Rotation of an electrolyte was observed around the direction of an external magnetic field both near the magnetic poles and the equator of the ball by means of the tracer particles as schematically shown in figure~\ref{figElRotat}. The opposite directions of rotation were observed for the steel and for the zinc-plated surfaces of a steel ball. The change of the direction of an external magnetic field into the opposite one resulted in the change of the direction of rotation of electrolyte in the both cases. The rotation of an electrolyte can be explained by the Lorentz force action on the nonuniform currents in an electrolyte. The currents in electrolyte are the manifestation of the magnetic field induced electric cell voltage. The latter arises due to the concentration difference of magnions in an electrolyte at the surface of a magnetized steel ball. As a result, the direction of rotation of an electrolyte is defined by the direction of an external magnetic field and the sign of the difference of electric field potential between the poles and the equator of the ball.

	The corrosion model is usually used to describe and model the processes of deposition (cementation) of metals without magnetic field application \cite{NowakHydromet84, MondalMatLet06, ShinagawaElchemSolSt09}. The deposition of copper or silver at iron or at zinc is considered as the work of a great number of chaotically distributed short-circuited galvanic cells where, according to this theory, iron or zinc dissolves at anodic regions, while copper or silver deposits at cathode regions, correspondingly. However, the current density of short-circuited galvanic cells decays at distances from the electrode about the diffusion layer thickness \cite{NowakHydromet84, MondalMatLet06} and it does not directly cause the stirring of an electrolyte at mesoscales (figure~\ref{figElRotat}) under the magnetic field effect.

\section{Discussion}
	
	The energy of an ion in an inhomogeneous magnetic field at the surface of a magnetized ferromagnetic electrode can be written as follows:
\begin{equation}
\Delta W_{\mathrm{m}}=-\frac{1}{2}\chi V \vec{H}^2 \left(\vec{r}_{\textrm{surf}}\right),
\label{MagnEnergy}
\end{equation}
where $\vec{r}_{\textrm{surf}}$ is the radius vector, whose value is set at the electrode surface, $\chi$ is the effective magnetic susceptibility of a magnion; $V$ is the magnion volume; $\vec{H}=\vec{H}_0+\vec{H}_{\mathrm{m}}$ is the magnetic field strength in the electrolyte which is the vector sum of the external homogeneous magnetic field strength $\vec{H}_{0}$ and the magnetic field strength created by the magnetized ferromagnetic electrode $\vec{H}_{\mathrm{m}}$. In particular, in the case of a ball-shaped electrode, the energy of an ion $\Delta W_{\mathrm{m}}$ can be written as $\Delta W_{\mathrm{m}}=-{1}/{2}\chi V \vec{H}^2\left(R,\theta\right)$, where $R$ is the ball radius, $\theta$ is an angle between the direction of an external magnetic field and the radius vector $\vec{r}$, and $\vec{H}_{\mathrm{m}}$ is the dipole magnetic field strength of a magnetized steel ball $\vec{H}_{\mathrm{m}} \left(r\right)={\left[3\left(\vec{m}\vec{r}\right)\vec{r}-\vec{m}r^{2} \right]}{r^{-5}}$, where $m$ is a magnetic moment of a steel ball.

	Let us use the standard approach of calculating the electromotive force of a concentration circuit to calculate the electric cell voltage of the circuit at the surface of a magnetized steel ball \cite{Bard01, Buchanan00, Vetter61, Thirsk74, Antropov72}.

	Moreover, let us write the distribution of concentration of magnions in an electrolyte in the vicinity of a magnetized steel electrode similarly to the references \cite{Bard01, Buchanan00, Vetter61, Thirsk74, Antropov72}:

\begin{equation}
C(\vec{r})=C_0\exp\left\{\frac{\chi_0\vec{H}^{2}\left(\vec{r}\right)}{2k_{\textrm{B}}T} \right\},
\label{Concentr}
\end{equation}
where $C(\vec{r})$ is concentration of magnions, $C_{0}$ is a constant, $\chi_0=\chi V$. The coordinate system origin coincides with the center of the magnetized steel ball.

	In particular, in the case of a magnetized electrode in the form of a ball, the square of the magnetic field strength $\vec{H}^{2}(\vec{r})=\vec{H}^{2}(r,\theta)$ in the expression (\ref{Concentr}) can be represented as follows:

\begin{equation}
\vec{H}^{2}\left(r,\theta \right)=M_{0}^{2} \left\{\xi ^{2} +\frac{8\pi \xi \left(3\cos ^{2} \theta -1\right)}{3} \left(\frac{R}{r} \right)^{3} +\frac{16\pi ^{2} \left(3\cos ^{2} \theta +1\right)}{9} \left(\frac{R}{r} \right)^{6} \right\},
\label{MagnField}
\end{equation}
where $M_{0}$ is the steel ball magnetization, $R$ is the steel ball radius, $\xi =H_{0}\left/M_{0}\right.$.

	It also follows from the general equation for electric cell voltage \cite{Bard01, Buchanan00, Vetter61, Thirsk74, Antropov72} that the electromotive force of concentration circuit between the points (we denote them 1 and 2) at the magnetized electrode surface is defined by the expression:
\begin{equation}
E_{12} =\frac{k_{\textrm{B}}T}{{Ze}}\ln\frac{a_{(2)}}{a_{(1)}}\,,
\label{EMF}
\end{equation}
where $Ze$ is the charge of the paramagnetic corrosion product, $a_{(1)}$, $a_{(2)}$ are the activities of ions at the points 1 and 2, correspondingly, at the steel ball surface. It is possible to transform the expression~(\ref{EMF}) for the electric cell voltage between the points at the surface of the ball-shaped electrode, which are at angles $\theta$ and $\theta=0$ relative to the direction of the external magnetic field $\vec{H}_0$, considering that the activity is approximately equal to the concentration at low values of  concentration \cite{Bard01, Buchanan00, Vetter61, Thirsk74, Antropov72}:
\begin{equation}
\varphi_{0\theta} =\frac{\chi _{0} }{2{Ze}} \left[\vec{H}^{2}(R,\theta)-\vec{H}^{2} (R,0)\right].
\label{Voltage}
\end{equation}

	Let us find the distribution of electric potential $\varphi$ and current density in the electrolyte volume. Continuity of the current density $\vec{j}$ is taken into account for this purpose:
\begin{equation}
\textrm{div}\vec{j}=0.
\label{Divergence}
\end{equation}

	The magnetic Reynolds number $R_{\mathrm{m}}={4\pi \sigma ul}/{c^{2}}$ is much less than unity $R_{\mathrm{m}} \ll 1$ because the electrolyte is a weekly conductive fluid. In this case, the current density can be represented as:

\begin{equation}
\vec j=\sigma\left(-\nabla\varphi+\frac{1}{c} \left[\vec{v}\times\vec{H}\right]\right),
\label{Current}
\end{equation}
 where $\sigma$ is the specific conductivity of an electrolyte, $l$ and $u$ are the characteristic parameters of the system size and fluid velocity, correspondingly \cite{FahidyElActa73, FahidyApplElChem83, AaboubiElChemSoc90}.
$\varphi$ is the electric field potential in the last formula, $\vec{v}$ is fluid velocity, $c$ is the velocity of light.

	Let us estimate the ratio
\[
\dfrac{\left|\dfrac{1}{c} \left[\vec{v}\times\vec{H}\right]\right|}{\left|\nabla \varphi \right|}
  \]
  to find the electric potential utilizing the experimental data \cite{IlchenkoJMMM10} where the external magnetic field \linebreak strength is equal to $H\simeq 1\div4$~kOe, velocity $v\simeq 2\div3$~cm~s$^{-1}$  and the voltage between the poles and the equator of a magnetized steel ball in electrolyte $\Delta \varphi \simeq 4\cdot 10^{-4}$~V. The ratio is about $10^{-1}$ to $10^{-2}$ in this case. That is why the problem solution can be found in the first approximation neglecting the second term in the formula (\ref{Current}) just for illustration. However, the solution qualitatively describes qualitatively the experimental picture for much greater magnetic field strength values and for the much greater values of the electrolyte velocities, correspondingly.

	Let us note that the specific electrical conductivity $\sigma $ of the electrolyte slightly varies with an increase or decrease of  concentration of the products of chemical reactions, and, therefore, may be a function of coordinates \cite{Antropov72}. Since the concentration of clusters of paramagnetic ions is small, we may neglect the dependence of specific conductivity of the electrolyte on the concentration of clusters and regarded it as a constant. Thus, after substituting the expression (\ref{Current}) in the equation (\ref{Divergence}) we obtain:
\begin{equation}
\Delta \varphi =0.
\label{LaplaseEq}
\end{equation}

	The solution of the equation (\ref{LaplaseEq}) has the form:
\begin{equation}
\varphi \left(r,\theta \right)=\frac{\chi _{0} }{2{Ze}} \left[\left(\frac{16\pi ^{2} M_{0}^{2} }{9} +\frac{8\pi M_{0} H_{0} }{3} \right)\left(3\cos ^{2} \theta -1\right)\left(\frac{R}{r} \right)^{3} +\frac{32\pi ^{2} M_{0}^{2} }{9} +H_{0}^{2} \right]
\label{Potential}
\end{equation}
and satisfies the boundary conditions:
\begin{equation}
\left\{\begin{array}{l} {\varphi \left(R,\theta \right)-\varphi \left(R,0\right)=\varphi _{0\theta}\,,} \\[1ex]
{\mathop{\lim }\limits_{\vec{r}\to \infty} \oint\limits_{S}\vec{j}\rd\vec{S}=0.}
\end{array}\right.
\label{Boundary}
\end{equation}

	The boundary conditions (\ref{Boundary}) take into account the coordinate dependence of the electric potential at the steel ball surface in the form (\ref{Voltage}) and conservation of the cumulative charge of the system.

	The resulting current density distribution is:
\begin{equation}
\vec{j}(r,\theta)=\frac{\sigma \chi _{0} }{2{Ze}} \left(\frac{16\pi ^{2} M_{0}^{2} }{3} +8\pi M_{0} H_{0} \right)\left[\left(3\cos ^{2} \theta -1\right)\frac{R^{3} }{r^{4} } \vec{e}_{r} +\sin 2\theta \frac{R^{3} }{r^{4} } \vec{e}_{\theta } \right],
\label{CurrentDensity}
\end{equation}
 where $\vec{e}_{r}$ and $\vec{e}_{\theta }$ are unit vectors of a spherical coordinate system.

\begin{figure}[!b]
\centerline{\includegraphics[width=0.6\textwidth]{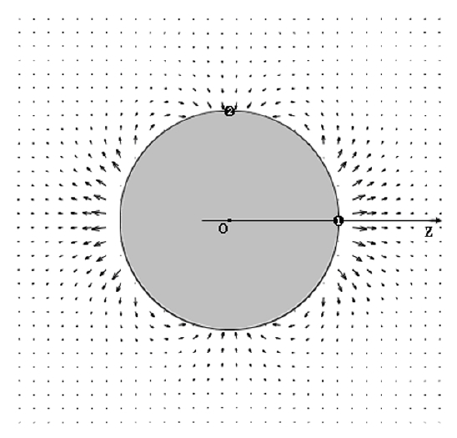}}
\caption{Scheme of the electric current density at etching a magnetized steel ball and at deposition of copper or silver at the magnetized steel ball. The external magnetic field and magnetization are directed horizontally and the angle $\theta $ is counted from the horizontal axis. The point 1 corresponds to the pole ($\theta =0$), the point 2 --- to the equator ($\theta =\pi\left/2\right.)$.}
\label{figCurrentDens}
\end{figure}
	The scheme of electric current directions in electrolyte is represented in figure~\ref{figCurrentDens} which arises while etching a magnetized steel ball in diamagnetic acid solutions and at deposition of copper or silver at the magnetized steel ball where the current density was calculated using the formula (\ref{CurrentDensity}). The electric current flows from the pole regions of the ball surface ($\theta =0$) to the equator regions ($\theta =\pi\left/2\right.$) as we see in figure~\ref{figCurrentDens}. The effectively paramagnetic reaction products move under the action of the gradient magnetic force from the equator to the poles of the ball creating in these cases a nonuniform concentration distribution (\ref{Concentr}) and the electric cell voltage. Accumulation of the effectively paramagnetic magnions in the vicinity of the poles of a magnetized steel ball is in accordance with the experimental results of this work.

\begin{figure}[!b]
\centerline{\includegraphics[width=0.4\textwidth]{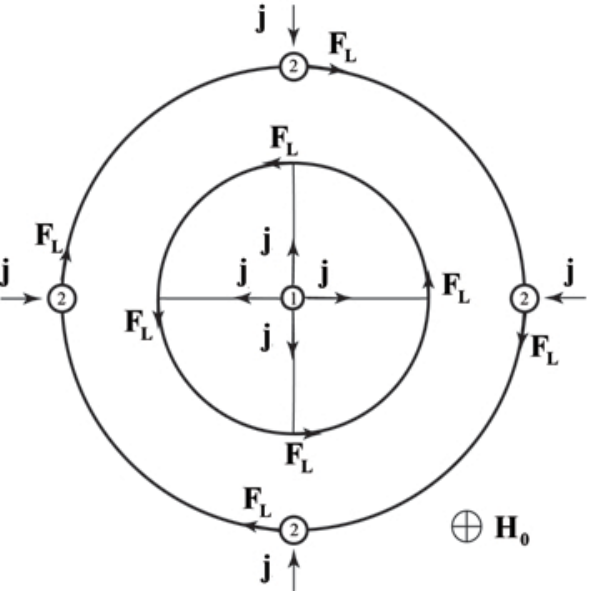}}
\caption{Scheme of the vortex flows of electrolyte arising near the poles and near the equator caused by the Lorentz force at etching of a magnetized steel ball or at deposition of copper or silver at a magnetized steel ball. The sight is chosen from the side of the pole. The sign $\oplus $ near vector $\vec{H}$ means that the external magnetic field is directed perpendicular to the plane of the picture from the observer.}
\label{figVortexFlows}
\end{figure}

If the sight is chosen from the side of a pole, then the current density will flow from the pole in an electrolyte (point 1, figure~\ref{figVortexFlows}). The scheme of the flows of the current density is represented in an electrolyte near the North magnetic pole in figure~\ref{figVortexFlows}. The Lorentz force $\vec F_{\textrm{L}}$ acts on the current density in the vicinity of the poles which leads to a rotation of an electrolyte around the direction of an external magnetic field. It is worth noting that the direction of electric current in an electrolyte is in accordance with the experiments in which rotation of an electrolyte is observed under Lorentz force action at etching of a magnetized steel ball or at deposition of copper or silver at a magnetized steel ball. The current density in an electrolyte flows into the equator (point 2, figure~\ref{figVortexFlows}). The Lorentz force acting on the current density near the equator leads to a rotation of an electrolyte (figure~\ref{figVortexFlows}) around the direction of the applied magnetic field but opposite to the direction of rotation near the poles. Exactly such a picture is observed in the experiments (figure~\ref{figElRotat}). The change of the direction of an external magnetic field $\vec{H}_0$ to the opposite one results in the change of the direction of rotation of an electrolyte both near the poles and the equator into the opposite ones. The directions of rotation of an electrolyte opposite to the ones represented in figure~\ref{figElRotat} are observed in the experiments of copper deposition at the zinc-plated surface of a magnetized steel ball as it was mentioned above. This is in accord with the formula (\ref{CurrentDensity}) in which the current density vector changes the sign at the change of the sign of the effective magnetic susceptibility.

It is possible to calculate the Lorenz force per unit volume of an electrolyte $\vec{F}_{\textrm{L}} ={c^{-1}}[\vec{j}\times\vec{H}]$ on the basis of the expression (\ref{CurrentDensity}):
\begin{equation}
\vec{F}_{\textrm{L}}=-\frac{4\pi\sigma\chi_0M_0^3R^3}{{Ze}cr^4}
\left(\frac{2\pi}{3}+\frac{H_0}{M_0}\right)\sin\theta\left[\frac{H_{0} }{M_{0} } \left(5\cos ^{2} \theta -1\right)+\frac{4\pi }{3} \frac{R^{3} }{r^{3} } \left(\cos ^{2} \theta +1\right)\right]\vec{e}_{\alpha }\,,
\label{LorenzForce}
\end{equation}
 where $\vec{e}_{\alpha }$  is the unit vector of the spherical coordinate system, $\alpha$  is the azimuth angle.

\begin{figure}[!b]
\centerline{\includegraphics[width=0.45\textwidth]{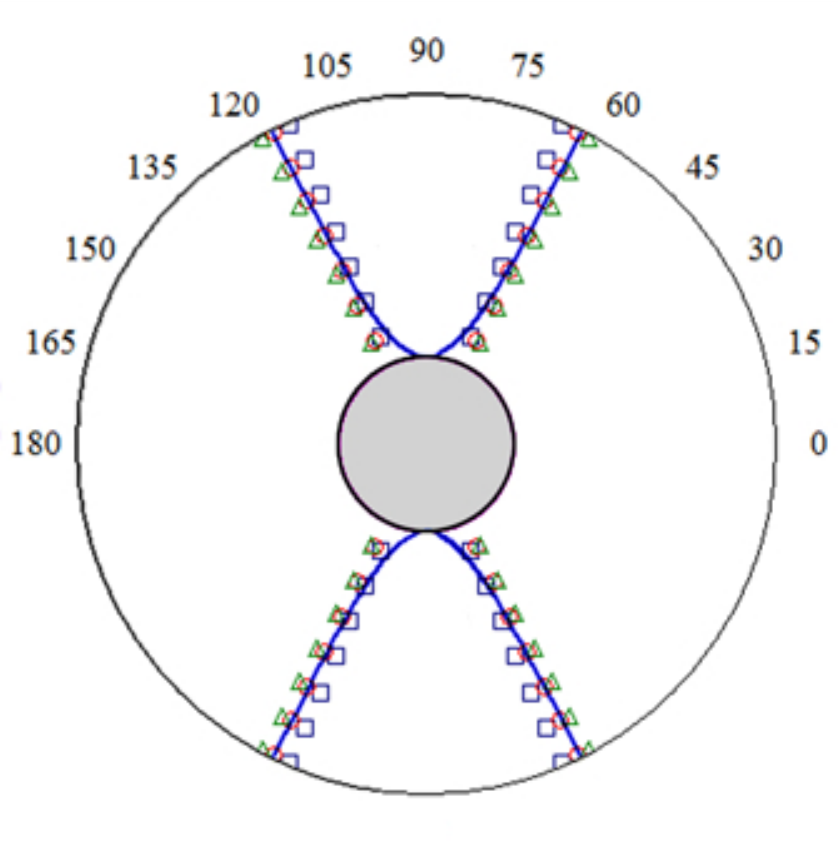}}
\caption{(Color online) The plot of the function $r(\theta)\left/R\right.$, the continuous curve is calculated by the formula (\ref{Surface}) at $H_0=7000$~Oe and ${4\pi M_0}/{3H_0}=1$. The figure shows experimental interfaces for the 1\% solution HNO$_3$ (curve is depicted by squares), 10\% solution FeCl$_3$ (circles) and 5\% solution CuSO$_4$ (triangles). Radius of the ball is equal to $R=2$~mm. Measurement error is 3\%.  The angle $\theta$ is counted in grades from the horizontal axis.}
\label{figSurf}
\end{figure}

The Lorentz force has only the rotational component which leads to a rotation of an electrolyte around the direction of an external magnetic field as it is seen from the expression (\ref{LorenzForce}). This is in accord with the experimental situation at the condition
\[
\dfrac{\left|\dfrac{1}{c} \left[\vec{v}\times\vec{H}\right]\right|}{\left|\nabla \varphi \right|}\ll 1.
\]
Besides, there is a cylindrically symmetric surface which is defined by the condition $\vec{F}_{\textrm{L}}=0$ at $r(\theta)>R$

\begin{equation}
r\left(\theta \right)=R\cdot \sqrt[{3}]{\frac{4\pi M_{0} }{3H_{0} } \frac{\left(1+\cos ^{2} \theta \right)}{\left(1-5\cos ^{2} \theta \right)}}\,,
\label{Surface}
\end{equation}
if ${4\pi M_{0} }/{3H_{0} } <1$, then the angle $\theta $ varies in the ranges of
$\theta _{\mathrm{cr}} < \theta \leqslant \theta _{0} $ and $\pi -\theta _{0} \leqslant \theta <\pi -\theta _{\mathrm{cr}}$,
where $\theta _{\mathrm{cr}} =\arccos \left(1/\sqrt{5}\right)={63.435}^{\circ } $, $\theta _{0} =\arccos \left(\sqrt{1-{4\pi M_{0}}/{3H_{0}}}\big/\sqrt{5+{4\pi M_{0}}/{3H_0}}\right)$, and if ${4\pi M_{0} }/{3H_{0} } \geqslant 1$, then the angle $\theta $ varies in the range of $\theta _{\mathrm{cr}} <\theta <\pi -\theta _{\mathrm{cr}} $. This surface separates the electrolyte regions with the opposite directions of rotation which is in accordance with the experimental data. If ${4\pi M_{0} }/{3H_{0} } >1$, then the surface  $r\left(\theta \right)$ does not intersect the ball surface and if ${4\pi M_{0} }/{3H_{0} } <1$, then it intersects the ball surface at $\theta =\theta _{0} $ and $\theta =\pi -\theta _{0} $. The plot of the function $r(\theta)$ is represented in figure~\ref{figSurf} at ${4\pi M_{0} }/{3H_{0} } =1$. Figure~\ref{figSurf} also presents experimental curves for the interface regions with opposite directions of rotation of various solutions near the surface of a magnetized sphere.

	Let us substitute the potential difference $\Delta \varphi \simeq 0.4\cdot 10^{-3}$~V between the pole ($\theta =0$) and the equator ($\theta =\pi\left/2\right.$) of a magnetized steel ball from the experimental work \cite{IlchenkoJMMM10} into the first formula of the equations (\ref{Boundary}) to estimate the ratio of the effective magnetic susceptibility of magnions $\chi _{0} $ to their charge $Ze$. The ratio $\chi_{0}\left/{Ze}\right.$ is at least three orders of magnitude bigger than such a ratio for the single Fe$^{2+}$ ion with the charge $2e$ and magnetic susceptibility defined by the Curie--Brillouin's law. As a result, it is possible to conclude that really a certain part of paramagnetic ions in an electrolyte do not have a free state but rather a bounded one in the form of magnions, similarly to the conclusions of the work \cite{FujiwaraPhysChemB04, ChiePhysChemB03}. The major contribution into the formation of magnions can be made by the  coacervates \cite{JongKolloid30, BamfordTrFaraday50, OverbeekCellCompPhys57, RozenblatMicroencaps89}, nanobubbles stabilized by the ions of an electrolyte \cite{KimColInter00, AttardPhysA02, IshidaInoueLangmuir00, TyrrellLangmuir02, ConsidineLangmuir99, MahnkePhysChem99, IshidaSakamotoLangmuir00, YakubovPhysChemB00} and by dispersive colloid nanoparticles surrounded by the ``clouds'' of ions that are widespread products of the grain-boundary corrosion \cite{WangElActa07, BennettMetTrans87, KellyElChemSoc93}. Moreover, the magnetic susceptibility of a single magnion in an electrolyte should be proportional to the quantity of ions in a group $n_\textrm{g}$ in the formula (\ref{Potential}). Then, the ratio ${n_\textrm{g}}{\left({Ze}/e\right)^{-1}}$ can be estimated as~$10^{3}$.

\section{Conclusions}

	In this work, the electromotive force of a concentration circuit and the current density distribution are calculated theoretically in an electrolyte at the processes of corrosion, chemical etching and deposition of metals under a nonuniform magnetic field choosing the uniformly magnetized steel ball under a uniform external magnetic field as an example and taking into account an inhomogeneous magnetic field in an electrolyte.

	First of all, it was shown in this work that a certain part of paramagnetic ions in an electrolyte in a magnetic field represent nanoclusters of effectively paramagnetic ions~--- magnions. The the coacervates \cite{JongKolloid30, BamfordTrFaraday50, OverbeekCellCompPhys57, RozenblatMicroencaps89}, micro- and/or nanobubbles stabilized by paramagnetic or diamagnetic ions in electrolytes  \cite{KimColInter00, AttardPhysA02, IshidaInoueLangmuir00, TyrrellLangmuir02, ConsidineLangmuir99, MahnkePhysChem99, IshidaSakamotoLangmuir00, YakubovPhysChemB00}, and colloid particles with their ionic surrounding can give a contribution into the formation of magnions. In this work, a significant difference of concentrations of reagents and reaction products arises between different regions of the steel ball surface due to the effect of a gradient magnetic force on magnions in inhomogeneous distribution of magnetostatic fields at its magnetization in an external magnetic field. The direction of a gradient magnetic force is defined by the sign of the effective magnetic susceptibility of magnions. The last can be changed, for example, by adding electrochemically inert paramagnetic ions or colloid particles to the solution, whereupon the way the direct and inverse deposition effects are observed \cite{DunnePhysRewB12, DunnePhysRewLett11, DunneMGD12, DunneApplPhys12}.  Summarizing the results of this work it is obvious that the inhomogeneous distribution of a magnetostatic field at the surface of both ferromagnetic and non-ferromagnetic electrode in an electrolyte and accumulation of the reaction products in the form of effectively para- or diamagnetic magnions lead to the creation of a magnetically induced electric cell voltage. This result is not restricted to only ball-shaped electrode but it is of a general nature taking into account the distribution of physical magnetic states at the electrode surface.

	The existence of magnions in an electrolyte has fundamental consequences for such a field of research as magnetoelectrolysis. The basic equations of magnetoelectrolysis such as equations of magnetohydrodynamics and convective diffusion should be supplemented by taking into account the gradient magnetic force acting on magnions. The boundary conditions for an electric potential should be changed in these equations taking into account the concentration electric cell voltage in a nonuniform magnetic field.  Material equations for these equations should also take into account the physical properties of magnions.

	The theoretical results of this work describe the experimental effects of self-organized electric current in an electrolyte from the magnetic poles, the equator of a ball or from the equator to the magnetic poles depending on the sign of an effective magnetic susceptibility of magnions, rotation of an electrolyte around the direction of an external magnetic field in the opposite direction near the poles and near the equator. The results of this work also describe the shape of the interface between the regions of an electrolyte with the opposite directions of rotation at the distances about the ball radius. The model of this work is valid at all stages of electrochemical transformations, and not only initially, until the form of the etching figure or the form of a deposit differs little from a spherical form. To take into account the changes of the form of the etching figure, it is necessary to substitute a particular distribution of magnetostatic stray fields in the electrolyte in general equations (\ref{Concentr})--(\ref{EMF}) instead of (\ref{MagnField}), in particular, for the case of a great lengthening of the ferromagnetic etching figure along the direction of the external magnetic field. These results represent a new way of controlling the shape of etching figures and a deposit structure which has an  important practical application.

	The effect of the magnetically induced electric cell voltage  arising in a magnetized steel ball differs in its nature from the electromotive force of the gravitational circuit \cite{ColliAnnPhysChem76} because the magnetic energy of the atoms inside the electrode volume is the same for a uniformly magnetized ball. It is necessary to note that the nature of the electric cell voltage arising between the surface regions in the magnetized ferromagnetic electrode can be more complicated in a general case than in the case of this work because if the shape of ferromagnetic electrode is different from the ellipsoid or if it is magnetized nonuniformly, then the distribution of magnetostatic field is nonuniform inside the electrode and thus different regions of the electrode material would have different physical conditions.

	Summing up the results of this investigation, we note the following:
\begin{itemize}
\item
	Nonuniform magnetic field exerts the greatest effect on the cluster components of an electrolyte, and, for rather large clusters, their energy in the magnetic field may exceed the energy of thermal motion, which leads to phase separation \cite{GorobetsJMMM13}, i.e., to the formation of a quasi-stationary heterogeneous state of an electrolyte and an inhomogeneous distribution of concentration of cluster components of an electrolyte on the electrode surface. In this case, the characteristic dimensions of the created phases (i.e., regions) with an increased concentration of cluster components relative to the rest of the electrolyte can be several orders of magnitude bigger than the characteristic thickness of the diffusion layer (figure~\ref{figVisual}, figure~\ref{figElRotat}), and thus similar effects are not related to the effects described in the papers \cite{DunnePhysRewB12, DunnePhysRewLett11, DunneMGD12, DunneApplPhys12} associated with deformation of the diffusion layer in an inhomogeneous magnetic field.	
\item
	Their effective magnetic susceptibility ``works'' in the presence of cluster components of the electrolyte in a nonuniform magnetic field that is equal to the difference of magnetic susceptibility of the cluster components and the fluid. When the magnetic susceptibility of an electrolyte changes from  diamagnetic to paramagnetic (by choice of different electrolytes), the sign of the effective magnetic susceptibility may also vary. This leads to the change of the location of areas of faster rate of electrochemical reactions at the electrode surface from the area with maximum magnetostatic field strength onto the area with the minimum of the latter, which was observed in several experimental studies \cite{GorobetsMatSc07, GorobetsPhysChemC08, GorobetsPhysMet05, GorobetsApSurfSc05, GorobetsPhysSolC04, GorobetsMetalPhys06, SueptitzCorSc11, IlchenkoJMMM10, GorobetsPhysMet12, GorobetsFuncMat11, DunnePhysRewB12, DunnePhysRewLett11, DunneMGD12, DunneApplPhys12, TschulikElCommun11, YangPhysChemLett12, UhlemannEurPhys13}. In the absence of clusters, i.e., for individual ions, it is impossible to introduce the concept of effective susceptibility since the latter is a thermodynamic quantity.
\item
	The formation of inhomogeneous concentration distribution of cluster components of an electrolyte on the electrode surface under the effect of an inhomogeneous magnetic field leads to the creation of an electrochemical concentration circuit, i.e., to the appearance of the electromotive force and the electrical current flow in an electrolyte between the electrode surface regions with different magnetostatic field strength. The formation of the regions of elevated concentration of paramagnetic ions but of another shape was also observed in the paper \cite{PullinsChemB01} in the vicinity of the magnetized iron microelectrodes (disks and cylinders) at transmitting electric current. The formed regions existed even after switching the electric current \cite{PullinsChemB01}. This seems possible due to the formation of magnions in these conditions as well.
\item
	Lorentz force arises due to the effect of magnetostatic field on the above described electric current in the electrolyte and induces the corresponding MHD stirring.  This effect was observed in several experimental studies \cite{IlchenkoJMMM10, GorobetsPhysMet12, GorobetsFuncMat11, DunnePhysRewB12, DunnePhysRewLett11, DunneMGD12, DunneApplPhys12}, and it is shown for the ferromagnetic electrode in the form of a ball in figures~\ref{figElRotat}--\ref{figVortexFlows}.
\end{itemize}

	The results of the experiments and the theoretical modelling can be used to create functional materials by means of magnetoelectrolysis and to model the effect of biogenic magnetic nanoparticles \cite{BlakemoreSc75, DunnBrainResBull95, SchultheissBioMetals97, DobsonFEBSLet01, KobayashiJapSocPowd97} on transport processes and biochemical reactions in cells of living organisms \cite{GorobetsNanoSc11}.

%
%
%

%
%% If you have problems with typesetting in ukrainian uncomment lines below.
%
%  \lastpage
%  \end{document}

\ukrainianpart

\title{Електрорушійна сила при травленні та осадженні металів \\ у неоднорідному постійному магнітному полі}
\author{О.Ю. Горобець\refaddr{label1,label2}, Ю.І. Горобець\refaddr{label1,label2}, В.П. Роспотнюк\refaddr{label1}, Ю.А. Легенький\refaddr{label3}}
\addresses{
\addr{label1} Національний технічний університет України ``КПІ'', просп. Перемоги, 37,  03056 Київ, Україна
\addr{label2} Інститут магнетизму НАН України та МОН України, просп. Вернадського, 36-б, 03142 Київ, Україна
\addr{label3} Донецький національний університет, вул. Краснозоренська, 40, Донецьк-87, Україна
}

\makeukrtitle

\begin{abstract}
\tolerance=3000%
Розраховано електрорушійну силу фізичного кола, що самоорганізовано виникає у зв'язку з неоднорідним розподілом концентрації ефективно діа- або парамагнітних компонентів електроліту на поверхні феромагнітного електроду під впливом неоднорідних магнітостатичних полів у процесах травлення та осадження металів на поверхні намагніченого феромагнітного електроду в електроліті. Також розраховано густину електричного струму та силу Лоренца в електроліті поблизу поверхні намагніченого сталевого електроду у формі кулі. Сила Лоренца спричиняє обертання електроліту навколо напрямку зовнішнього магнітного поля.
\keywords магнетоелекроліз, градієнтна магнітна сила, сила Лоренца, ефективна магнітна сприйнятливість, кластери

\end{abstract}

\end{document}